\documentclass[a4paper,fleqn,usenatbib]{mnras}

\usepackage{mathptmx}
\usepackage{textcomp}

\usepackage[T1]{fontenc}
\usepackage{ae,aecompl}

\usepackage{graphicx}	
\usepackage{amsmath}	
\usepackage{amssymb}	
\usepackage{pdflscape}	

\let\orgautoref\autoref
\renewcommand{\autoref}
        {\def\equationautorefname{Eq.}%
         \def\figureautorefname{Fig.}%
         \def\sectionautorefname{Sect.}%
         \def\subsectionautorefname{Sect.}%
         \def\subsubsectionautorefname{Sect.}%
         \orgautoref}

\title[Periodic variability of a $z$=2.0 quasar]{Periodic variability 
of the $z$=2.0 quasar QSO B1312+7837\thanks{Based on observations 
gathered at the Rozhen National Astronomical Observatory.}}

\author[M. Minev et al.]{
M. Minev,$^{1,2}$
V.D. Ivanov,$^{3}$
T. Trifonov,$^{4}$
E. Ovcharov,$^{1}$
S. Fabrika,$^{5}$
O. Sholukhova,$^{5}$
\newauthor
A. Vinokurov,$^{5}$
A. Valcheva,$^{1}$
P. Nedialkov$^{1}$
\\
$^{1}$Department of Astronomy, Faculty of Physics, University St. Kliment Ohridsky, 05 James Bourchier, Sofia, Bulgaria.\\
$^{2}$Institute of Astronomy and NAO, Bulgarian Academy of Sciences, 72 Tsarigradsko Shose Blvd., 1784 Sofia, Bulgaria\\
$^{3}$European Southern Observatory, Karl Schwarzschildstr. 2, D-85748 Garching bei M\"unchen, Germany\\
$^{4}$Max-Plank-Institut f\"ur Astronomie, K\"onigstuhl 17, Heidelberg 69117, Germany\\
$^{5}$Special Astrophysical Observatory of the Russian Academy of Science (SAO RAS), Nizhnij Arkhyz, Karachai-Cherkessia, Russia\\
}

\date{Accepted 2021 September 21. Received 2021 September 20; in original form 2021 January 17}

\pubyear{2021}

\begin{document}
\label{firstpage}
\pagerange{\pageref{firstpage}--\pageref{lastpage}}
\maketitle

\begin{abstract}
We report here the first results from a 15-yr long variability monitoring of the z=2.0 quasar QSO B1312+7837. It shows
luminosity changes with a period P$\sim$6.13 yr (P$\sim$2.04 yr at rest frame) and an amplitude of $\sim$0.2\,mag, superimposed on a
gradual dimming at a rate of $\sim$0.55\,mag per 100\,yrs. Two false periods associated with power peaks in the data
windowing function were discarded. The measured period is confirmed with a bootstrapping Monte-Carlo simulation. A damped
random walk model yields a better fit to the data than a sine-function model, but at the cost of employing some high
frequency variations which are typically not seen in quasars. We consider the possible mechanisms driving this variability,
and conclude that orbital motion of two supermassive black holes -- result from a recent galaxy merger -- is a possible
explanation. 
\end{abstract}




\section{Introduction}

Supermassive black holes (SMBHs) reside at the centers of massive 
galaxies \citep{1970ApJ...161..419W}. They dominate the kinematic 
evolution of the central regions of galaxies, and affect the evolution 
of their stellar populations. 
During the phases of active accretion, the AGN (Active Galactic Nucleus) phenomenon occurs, 
giving rise to the quasars, that are used to probe the distant Universe 
\citep{1975Natur.254..295H}.

Galaxies often interact with each other and their merging can form a 
new nucleus that contains two SMBHs \citep{1980Natur.287..307B}. The 
most direct search for binarity of SMBHs is to look for spatially 
resolved sources of X-ray, radio emission or broad line optical emission 
within the same host galaxy, but this is usually limited to nearby 
objects 
\citep[NGC\,6240, Mrk\,212;][]{2003ApJ...582L..15K,2021MNRAS.500.3908R}
and difficult, even by interferometry \citep{2018ApJ...863..185D}. 
However, some SMBHs can be ejected from the host galaxy -- 
\citet{2021ApJ...913..102W} reported nine candidates. 
Double-peak lines, associated with AGNs also indicate binarity 
\citep{1988Natur.331...46H,1994ApJS...90....1E,2010ApJ...716..866S,2021A&A...646A.153S}.
Another option -- adopted here -- is to look for periodicity of the 
emission from unresolved AGNs, modulated by the orbital motion of the 
two SMBHs 
\citep{1988ApJ...325..628S,1998ApJ...507..173F,2015MNRAS.453.1562G,2016ApJ...833....6L,2016MNRAS.463.2145C}. 
In an extensive review \citet{2006MmSAI..77..733K} lists a few other 
indicators of SMBH binary: double-double radio galaxies and X-shaped 
radio galaxies.

Various mechanisms that cause this variability have been considered. 
One possibility is that the orbital motion modulates the accretion 
rate on the SMBHs \citep[][]{2013MNRAS.436.2997D}. Another is a 
transit of the ``secondary'' SMBH in front of the accretion disk of 
the ``primary'' SMBH \citep[][]{1996ApJ...460..207L}. This occurs when 
the distances between SMBHs are less than 1\,pc, and in such cases 
the two components can spiral towards the common center of mass and 
eventually merge \citep{1980Natur.287..307B}. The orbital motion 
causes strong gravitational waves that should be detectable from the 
latest experiments \citep{2019MNRAS.485.1579K}, but such events are 
rare. The studies of SMBHs binaries are important for understanding 
the galaxy mergers, for nature of SMBHs themselves, and for the 
physics of the gravitational waves.

Here we report the results of a long-term variability monitoring of 
the quasar QSO\,B1312+7837 
\citep[\protect{$[$VV2006$]$ J131321.4+782153}, WISEA J131321.33+782153.8, Gaia DR2 1716672593984035072;][]{2006A&A...455..773V} at $z$=2.0 
\citep{1999A&AS..134..483H}. It shows the typical broad emission lines
\citep{2004AJ....128.1058T}. Its mean magnitude is $B\sim16.4$\,mag, 
corresponding to an absolute magnitude of $M_B$=$-$30.1\,mag 
\citep{1999Afz....42....5M}, $\sim$3.5\,mag brighter than the mean 
$M_B$ for quasars at the same redshift \citep{2015A&A...583A..75S}. 
Intranight observations with a minute-long cadence showed no short-term 
variability \citep{2005MNRAS.358..774B}.	


\section{Observations}\label{sec:observations}

\begin{figure}
\begin{center}
\includegraphics[width=8.4cm]{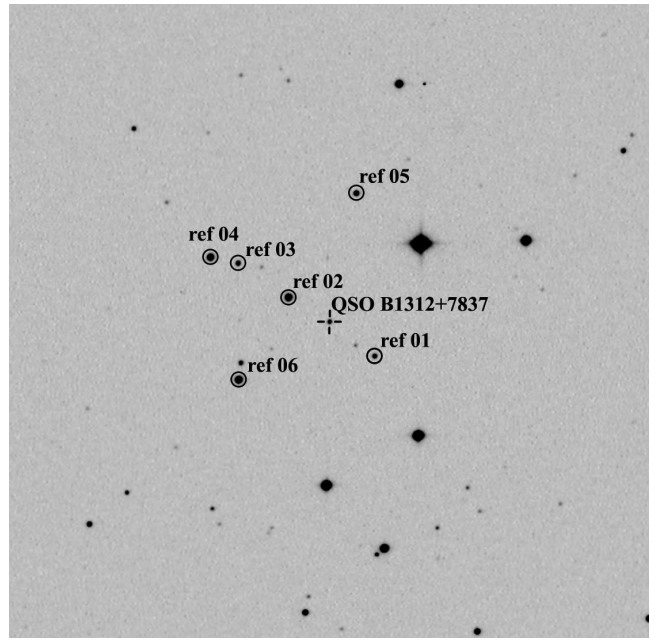}
\end{center}
\caption{The field (10$\times$10\,arcmin) around QSO B1312+7837 
(cross). The stars used as references are marked with circles. 
North is up, east is left.}
\label{fig:FoV}
\end{figure}

The observations were obtained at the Rozhen National Astronomical Observatory (NAO)
with a number of imagers \citep[Table\,\ref{tab:facilities} and for the 
FoReRo-2 -- in ][]{2000KFNTS...3...13J} 
equipped with standard photometric Johnsons-Cousins $UBVRI$ filters. The 
integration times were 1.5-5\,min and the seeing was 1.5-2.0\,arcsec.
The field around the quasar and the reference stars are shown in 
Fig.\,\ref{fig:FoV}. The observing campaign starts from JD\,2453494 and
the quasar's apparent $R$ band magnitudes are in range 15.89-16.20\,mag 
with typical photometric error 0.01-0.04\,mag. A sample of the photometric 
data of the quasar and two reference stars (ref-01 -- USNOA2\,1650-01631981 
and ref-05 -- USNOA2\,1650-01632068) is show in Table~\ref{tab:obslog} and 
the entire data is available in the electronic edition of the journal. 
There is no evidence for variability of these two stars 
\citep{2016yCat.2336....OH} and as we can see from our observations their 
magnitude is constant in time. Comparison between magnitudes of any of 
these stars and the quasar shows variability that cannot be produced by 
some correlated image noise or the different observational equipment.	

\begin{table}
\caption{Observing facilities. The columns contain: telescope, camera,
field of view (FoV) and pixel scale.}\label{tab:facilities}
\begin{center}
\begin{tabular}{@{}l@{ }l@{ }c@{ }c@{}}
\hline
Telescope & Camera & FoV,\,\arcmin & ~Scale,\,\arcsec\,px$^{-1}$ \\
\hline
50/70cm Schmidt & SBIG\,ST-8       & 27.5$\times$18.3 & 1.08 \\
50/70cm Schmidt & SBIG\,STL-11000M & 72.1$\times$48.1 & 1.08 \\
50/70cm Schmidt & FLI\,PL-16803    & 73.7$\times$73.7 & 1.08 \\
2m RCC          & VersArray\,1300B & 5.8$\times$5.6   & 0.26 \\
2m RCC          & FoReRo-2         & 17.1$\times$17.1 & 0.50 \\
\hline
\end{tabular}
\end{center}
\end{table}

The data were processed with the Image Reduction and Analysis 
Facility 
\citep[IRAF;][]{1986SPIE..627..733T,1993ASPC...52..173T}\footnote{IRAF 
is distributed by the National Optical 
Astronomy Observatory, which is operated by the Association of 
Universities for Research in Astronomy under a cooperative  
agreement with the National Science Foundation.} and included 
the usual steps: bias/dark subtraction, flat fielding, and flux 
calibration. Aperture photometry was performed using the 
{\it APPHOT} IRAF package. The images were flux-calibrated with 
standards from \citet{2000PASP..112..925S}. 
To exclude variable 
sources we considered only stars with rms\,$\leq$\,0.2\,mag. For 
the VersArray\,1300B data we applied an additional correction to 
remove a known radial flux variation \citep{2008MNRAS.386..819O}.

\begin{table}
\caption{Photometric light curve QSO\,B1312+7837. Only the first 
five entries are shown for guidance, the entire table is 
available in the electronic edition of the journal. The columns 
contain: Julian date, standard $R$ band magnitude and its error for the quasar and two reference stars.}
\label{tab:obslog}
\begin{center}
{\scriptsize
\begin{tabular}{ccccccc}
\hline
 JD$-$2453000 & QSO\,$R$ & err & ref-01\,$R$ & err & ref-05\,$R$ & err \\
 & mag & mag & mag & mag & mag & mag \\
\hline
 494.350 & 15.942 & 0.083 & 15.558 & 0.082 & 15.099 & 0.082 \\
 494.500 & 15.922 & 0.078 & 15.557 & 0.078 & 15.107 & 0.077 \\
 623.325 & 16.037 & 0.020 & 15.655 & 0.019 & 15.190 & 0.019 \\
 624.335 & 16.035 & 0.017 & 15.673 & 0.015 & 15.202 & 0.014 \\
 626.460 & 16.041 & 0.019 & 15.659 & 0.018 & 15.208 & 0.017 \\
\hline
\end{tabular}
\par}
\end{center}
\end{table}

\begin{figure}
\begin{center}
\includegraphics[width=8.4cm]{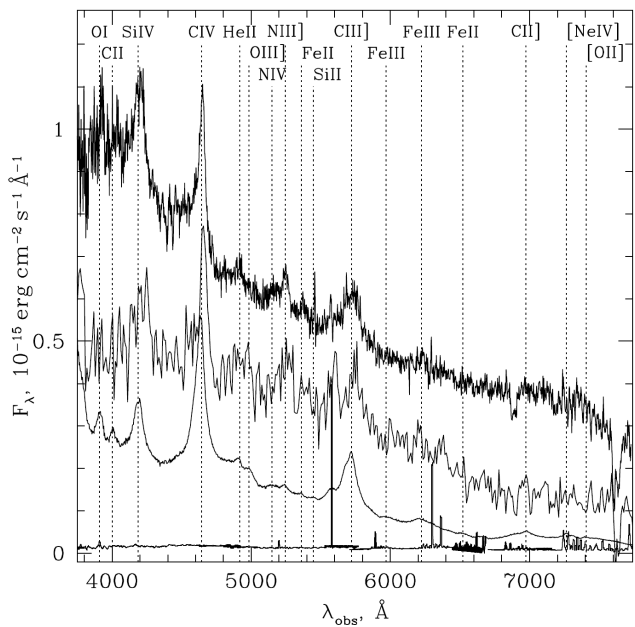}
\end{center}
\caption {Optical spectra at observed wavelengths of (from top to 
bottom): QSO B1312+7837 from the 6-m BTA telescope and from the 2.2-m 
telescope on Calar Alto \citep[shifted down by 0.3 for display 
purposes;][]{1999A&AS..134..483H}, a combined SDSS quasar spectrum 
shifted to $z$=2 \citep{2001AJ....122..549V} and a sky spectrum. 
Some prominent quasar emission features are marked.}
\label{fig:spectrum}
\end{figure}

A spectrum of QSO B1312+7837 was obtained on Feb 18, 2018 at the 
6-m BTA telescope at the Special Astrophysical Observatory with 
SCORPIO \citep[Spectral Camera with Optical Reducer for Photometric 
and Interferometrical Observations;][]{2005AstL...31..194A}, 
equipped with 2048$\times$2048 EEV\,42-40 detector, yielding a 
field of view of 6.1\,arcmin on the side. The slit was 1\,arcsec 
wide and the volume phase holographic grating VPHG550G was used, 
delivering a resolving power R$\approx$550 and FWHM = 10\,\AA\, 
over a wavelength range $\lambda 3700-7700\,$\AA . The integration 
time was 3000\,sec (5$\times$600\,sec exposures). 
The data processing included bias/dark subtraction, flat fielding, 
extraction, wavelength calibration and a flux calibration. The 
final spectrum is shown in Fig.\,\ref{fig:spectrum}.

We measured a redshift 1.999$\pm$0.004 for 
N\,{\sc v} 1240, Si\,{\sc iv} 1394, Si\,{\sc iv} 1403, C\,{\sc iv} 1549,
He\,{\sc ii} 1640, Ne\,{\sc iii} 1750, C\,{\sc iii}] 1909 lines.
It is indistinguishable from the value of \cite{1999A&AS..134..483H}.
The error was determined as in \citet{2016A&A...588A..93I}, as a
standard deviation of the redshifts measured from the seven lines -- 
with the caveat that the standard deviation for a small number of 
measurements is not well-define, but we based this estimate only on 
relatively strong and isolated lines, without apparent intervening 
absorptions.

\begin{figure}
\begin{center}
\includegraphics[width=9cm]{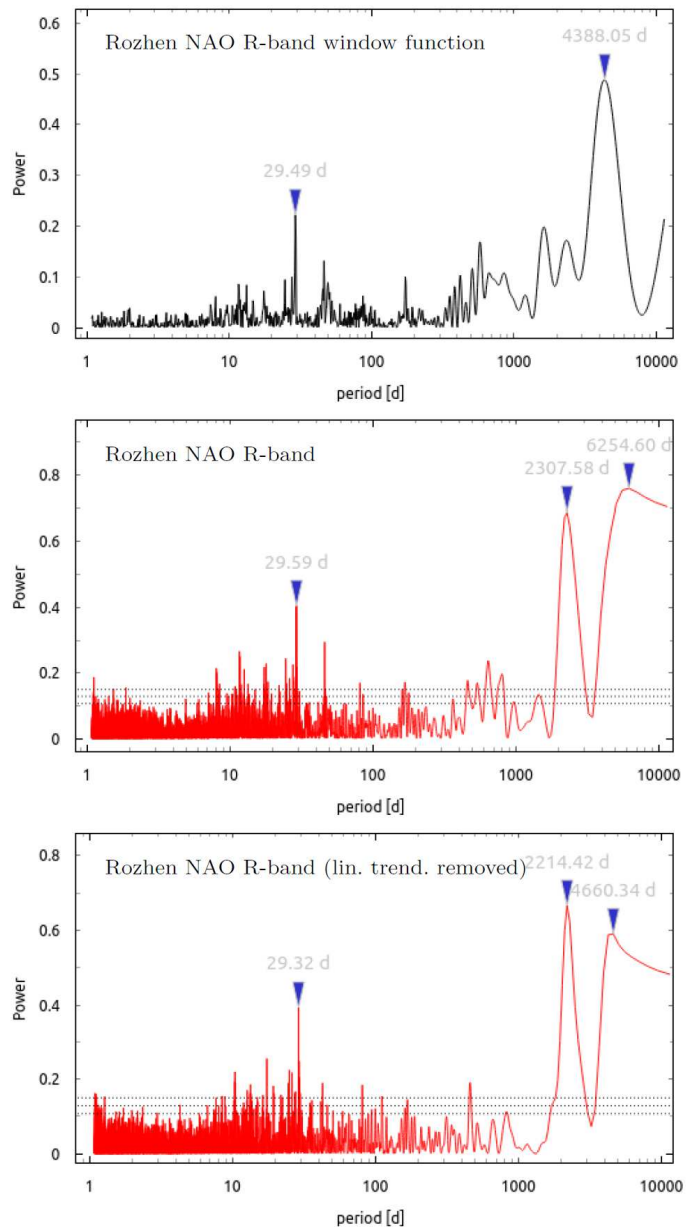} \\
\end{center}
\caption{ {\em Top panel}: Window function (WF) power spectrum of the Rozhen NAO R-band photometry for QSO B1312+7837. The observational scheduling induces strong WF power at periods of 29.49\,d, and 4388.05\,d, among others. {\em Middle panel}: GLS power spectra of the Rozhen NAO R-band. Horizontal dashed lines indicate the FAP thresholds of 10\%, 1\% and 0.1\%. The three most significant powers are at periods 29.59\,d, 2307.58\,d, and 6254.60\,d. {\em Bottom panel}: Same as the panel above, but after a linear trend model is subtracted from the R-band photometry. The true period is near 2214.42\,d, whereas the 4660.34\,d and 29.32\,d are its aliases of the WF peaks at periods of  4388.05\,d and 29.59\,d, respectively.}
\label{GLS} 
\end{figure}

\section{Analysis}\label{sec:analysis}

\subsection{Period search}

For quasar periodicity search, we used a modified version of the {\tt Exo-Striker} tool
\citep{2019ascl.soft06004T}, which provides easy access to a large variety of algorithms for time-series analysis. To search for periodic signals in the photometry, we first constructed a generalized Lomb-Scargle \citep[GLS;][]{2009A&A...496..577Z} power spectrum periodogram on the Rozhen NAO R-magnitude data, which have the longest temporal baseline and best quality. We adopted a false alarm probability (FAP) 
a significance threshold of 10$^{-3}$ (0.1\%), and we inspected the frequency range between one day and two times the length of the observations' temporal baseline, which is 5485 days. For our GLS periodogram tests, we always included a white-noise model. The variance is quadratically added to the error budget of the R-magnitude data.   
Additionally, we perform a Discrete Fast-Fourier analysis of the Rozhen NAO R-magnitude data to study the Window Function (WF) of the data, affecting our period search.

\autoref{GLS} shows the results from our WF and GLS periodicity search test.
The top panel of \autoref{GLS} shows WF power, which shows many peaks. The strongest is at 4388 d and there is another one
at 29.5 d that is close to the Lunar synodic-month, affecting the observational schedule.
The middle panel of \autoref{GLS} shows the GLS periodogram of the Rozhen NAO R-magnitude data.
We found two very significant GLS peaks: at a period of 2307.6\,d (FAP < 10$^{-49}$), and another at a period of 6254.6\,d (FAP < 10$^{-58}$) centered at a broad, low-frequency power structure. However, 
the latter has a longer period than the temporal baseline, thus it is not a firm detection. An adequate explanation of the significant low-frequency GLS power is the presence of a signal component, which cannot be resolved on our limited temporal baseline.
Indeed, a visual inspection of the all filters photometric data suggests a gradual magnitude decline. This motivated us to apply a 
linear trend fit to the NAO R-magnitude data, which confirmed the presence of a
significant photometric magnitude decline (see \autoref{Sec3.2}). 
The bottom panel of \autoref{GLS} shows the GLS periodogram of the best-fit residuals of the linear trend model. We still find two significant low-frequency signals. Compensating the brightness decay,
reduced the power and the period of the lowest-frequency signal, which is now detected near $\sim$ 4660\,d, whereas the second strongest peak in the raw R-band photometry is now the most significant peak, but with a slightly different period at 2214.4\,d. In addition,
we find evidence of a strong GLS period near $\sim$ 29\,d. 
We find, however, that the 4660.34\,d signal and the 29.32\,d signals are most likely related to the aliases of the WF and the true period of 2214.42\,d.
Indeed, P$_{\rm alias~1}$ = $1/(f_{\rm WF=29.5\,d} - f_{\rm 2214\,d}) \approx 29.6\,d.$ and  P$_{\rm alias~2}$ = $1/(f_{\rm 2214.4} - f_{\rm WF=4388\,d}) \approx 4500\,d.$
Therefore, we 
concluded that the periodic signal evident in the data is at a period near 2214\,d, whereas the remaining strong power in the GLS power is likely induced by a combination of an additional photometric variability of unknown nature that appears as a linear trend, and the WF aliasing with the true period.

\subsection{Parameter optimization and model selection.}
\label{Sec3.2}

We adopted the Simplex algorithm \citep{NelderMead65}, which 
optimizes the negative logarithm of the likelihood function 
($-\ln\mathcal{L}$), coupled with three competing models; 
(i) a ``null`` model assuming no signal in the data.  
(ii) A sinusoidal model where the optimized parameters are 
the photometric signal, amplitude, phase, and period. 
(iii) A sinusoidal model as (ii), but with an additional linear 
brightness decline term. 
For models (ii) and (iii), we adopted the 2214.4-day peak, 
phase, and amplitude estimate from the GLS analysis as an 
initial guess for parameter optimization.
Additionally, for all models we vary the nuisance parameters: 
the mean ``offset'' of the photometric data, and the white-noise variance
term, which we add in quadrature to the nominal R-band uncertanties while fitting \citep[i.e., a data ``jitter'' term, see][]{2009MNRAS.393..969B}. 
For posterior analysis, we ran an affine-invariant ensemble Markov 
Chain Monte Carlo (MCMC) sampler \citep{2010CAMCS...5...65G} via 
the \texttt{emcee} package \citep{2013PASP..125..306F}. 
We adopted non-informative flat priors of the parameters, and we 
explored the parameter space starting from the best-fit parameters 
returned by the Simplex minimization. We ran 100 independent 
walkers in parallel adopting 1000 ``burn-in'' MCMC steps, which we 
discard from the analysis, followed by 5000 MCMC steps, from which 
we constructed the posterior parameter distribution.
We evaluate the acceptance fraction of the \texttt{emcee} sampler, which 
as recommended by \citet[][]{2010CAMCS...5...65G}, should be between 0.2 and 0.5,
to consider whether the MCMC chains have converged.
We adopted the 68.3\% confidence intervals of the MCMC 
posterior distributions as a $1\sigma$ uncertainty estimate of the 
parameters.

\begin{figure}
\begin{center}
\includegraphics[width=8.cm]{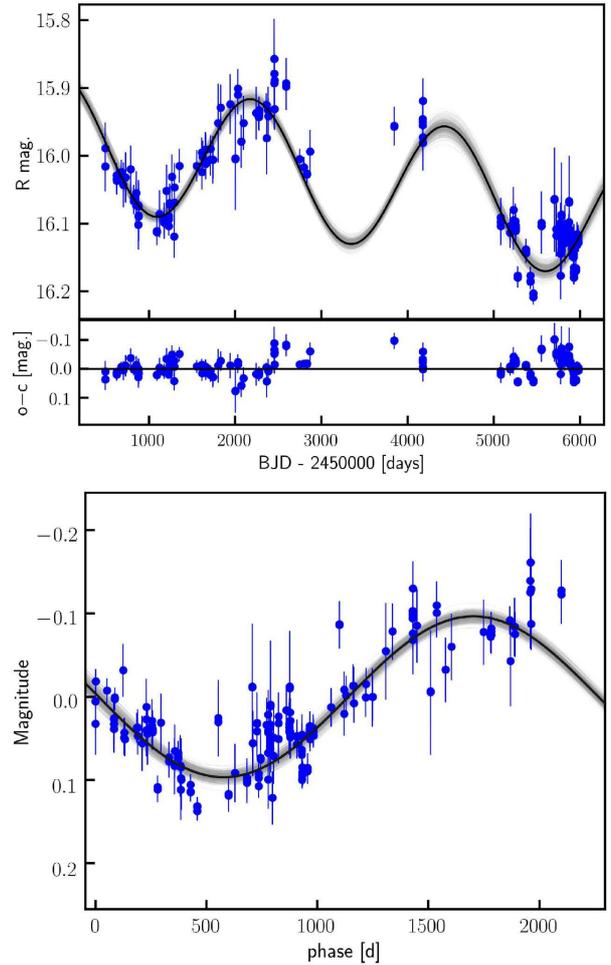}
\end{center}
\caption{The top panel shows the best {\tt Exo-Striker} fit (black) to the $R$ band 
 observed data (blue) on the first panel, and 
residuals on the second panel. The fit is composed of strictly periodic component and a brightnest decline term, which represents the data well.
The bottom panel shows a period-folded $R$ band light curve.
The shaded area is composed of 500 randomly chosen fits from the 
MCMC test, which visually represents the 1\,$\sigma$ uncertainties of the model.}
\label{fig:LC_ES}
\end{figure}

The best-fit statistical properties of three competing models 
in terms of Bayesian Information Criteria\footnote{$BIC$ is 
defined as $-$2$\times\ln\mathcal{L}$ + k$\times\ln$(N), where k 
is the number of free parameters in the model, and N is the number 
of data. Two competing models with different $k$ can be tested via 
their $\Delta BIC$, which must be over 10 to support the more 
complex model.} (BIC) are as follows: $BIC_{\rm flat}$ = $-$414.98, 
$BIC_{\rm Sine}$ = $-$618.25, and $BIC_{\rm Sine + trend}$ = $-$739.37.  
With a $\Delta BIC$ = $BIC_{\rm flat} - BIC_{\rm Sine + trend}$ = 324.39, 
represents a very strong evidence in support of the quasar periodic 
variability.
Our final results derived a quasar variability with a period of 
2237$\pm$12\,d (corresponding to 746$\pm$4\,d
at frame of rest). The light curve is shown in Fig.\,\ref{fig:LC_ES}.

As a consistency check we carried out an independent analysis fitting
the observational light curve with a combination of a 
sinusoidal and a linear trend. To estimate the errors of the derived 
parameters we carried out a semi-empirical Monte-Carlo bootstrapping 
simulation, following \citet{2009A&A...507..481C,2011A&A...530A...5C}:
after we obtain the best fit, for the i$^{th}$ observation we calculate
the deviation from the best fit and then we add this deviation to the 
(i+1)$^{th}$ observation. In a circular fashion, the deviation for the 
last observation is added to the first observation. Thus, we create a 
new sample and repeat the fit. In the next step we add the deviation 
for the i$^{th}$ observation to the (i+2)$^{th}$ observation, and 
re-fit, until we have created a number of realizations equal to the 
number of observations minus one. A drawback of this method is that 
the number of realizations is limited by the number of observations 
but the procedure preserves any 
systematic effects that may be present in the data: if there is a 
cluster of measurements with larger errors, e.g. because there was a 
batch of adjacent night with poor weather conditions, this structure 
will be preserved in the simulation and will contribute in the same 
systematic way to the uncertainties of the derived parameters. The 
results are listed in Table\,\ref{tab:resuls} and they agree, within 
the errors with the {\tt Exo-Striker} analysis.

\begin{table}
\caption{Results from the Monte-Carlo simulation.
\label{tab:resuls}}
\begin{center}
\begin{tabular}{@{ }l@{ }c@{ }c@{ }c@{ }}
\hline
Parameter & mean & median &~std. dev. \\
\hline
Sin Amplitude, mag          & 0.095    & 0.098    & 0.010    \\
Sin Period, days            & 2248.2   & 2253.3   & 53.1     \\
Sin Phase, days             & $-$1.45  & $-$1.43  & 0.28     \\
Linear Slope, mag$^{-1}$\,days &~1.52e-05~&~1.51e-05~&~0.40e-05 \\
Average magnitude, mag  & 15.977   & 15.974   & 0.014    \\
RMS, mag                & 0.0065   & 0.0062   & 0.0017   \\
\hline
\end{tabular}
\end{center}
\end{table}

\subsection{Stochastic variability and damped random walk model}

It has been suggested in the literature that quasars' optical variability could be well explained with red-noise stochastic processes instead of other physical phenomena such as 
close binary supermassive black hole \citep[see, e.g.,][]{2009ApJ...698..895K, 2010ApJ...708..927K, 2010ApJ...721.1014M}. 
In this context, it is essential to test the probability of the optical variability of QSO B1312+7837 being a true periodicity rather than a manifestation of correlated red noise. A commonly used stochastic process for modeling  optical variability in quasars is the damped random walk (DRW) model \citep{2009ApJ...698..895K}, which could be tested against our periodic best-fit presented in \autoref{Sec3.2}.

For the purpose we adopted the {\tt RealTerm} Gaussian process (GP) regression kernel intrinsic to the {\tt celerite} Python package \citep{2017AJ....154..220F}, which is included in the {\tt Exo-Striker}. By definition, this model is a damped random walk process of the form: $\kappa(\tau)=a_{j}e^{-c_{j}\tau}$, where $a$ is an amplitude and $c$ defines the characteristic timescale of the GP, therefore, suitable for our needs.
For consistency with our best-fit periodic model, the $a$ and $c$ GP hyper-parameters are fitted 
together the mean offset and the white-noise variance nuisance parameters of the R band data.  Thus, the DRW model, and our best fit periodic model are nested within the null model,which assumes no periodicity. 

For completeness, we also performed DRW fits with a linear trend component, and with a sine-model component, which includes a linear trend. Thus way, the DRW co-variance in the likelihood function allowed us to test if it is statistically worthy of the involvement of a liner, periodic, or both components beyond the possible correlated DRW variations. 

We achieved DRW parameters estimates of
$a$ = 0.0058$_{-0.0001}^{+0.0017}$, and  $c$ = 0.0015$_{-0.0011}^{+0.0002}$, leading to 
a low rms = 0.0117 mag, and $BIC_{\rm DRW}$ = $-$849.54.
With $\Delta BIC$ = $BIC_{\rm Sine + trend}$ - $BIC_{\rm DRW}$ = 110.17, this gives a significant advantage to the DRW model with respect to our adopted periodic model. 
The addition of a linear trend to the DRW model decreased the BIC evidence to $BIC_{\rm DRW+ trend}$ = $-$847.49, which means that assuming we observe stochastic processes in our data, the addition of this parameter is not well justified.  The inclusion of a periodic component to the latter model, however, gives a better $BIC_{\rm DRW + Sine + trend}$ = $-$853.22. This leads to positive evidence of $\Delta BIC$ = 3.68 with respect to the DRW-only model, which means that the NAO-Rozhen data of QSO B1312+7837 could indeed be consistent with a periodic behavior under the assumption that we predominantly observe correlated red-noise variations.

\begin{figure*}
\begin{center}
\includegraphics[width=17cm]{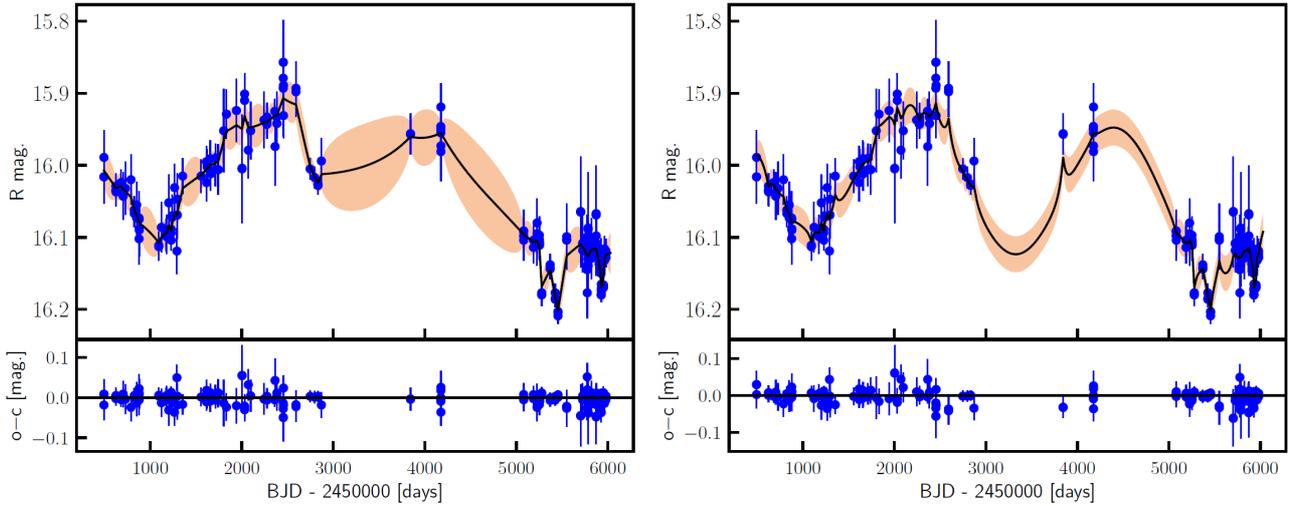}
\end{center}
\caption{Same as top panel of \autoref{fig:LC_ES}, but for an alterantive DRW GP model with a linear trend (left), and a DRW GP model with a linear trend and a periodic component (right). The balck solid curve shows the best fit model, whereas the computed GP covariance is shown with orange. The residuals are shown on the lower sub-panels.}
\label{fig:DRW_ES}
\end{figure*}

We concluded that the DRW + Sine + linear trend model is by far the best model, although only marginally better than a simpler DRW-only model. 
\autoref{fig:DRW_ES} shows the DRW best-fit models and their computed GP covariance
in addition to \autoref{fig:LC_ES}, which shows our best periodic model.  
Indeed, a visual inspection of the DRW residuals in \autoref{fig:DRW_ES} 
suggests nearly perfect agreement with the DRW models. Can then red-noise stochastic processes explain the optical variability of QSO B1312+7837?

While we cannot reject this possibility, we warn that our DRW model
has at least a few major caveats. First, the DRW model is consistent with high-frequency behavior, which has a strong tendency to overfit data. In this context, the DRW model would be adequate only if we have strong priors on the stochastic processes in quasars, which could constrain $a$ and $c$.
Second, a direct comparison between a simple strictly periodic Sine model composed of linear parameters\footnote{The period $P$ is a non-linear parameter, but is strongly constrained by the GLS test} and a highly flexible GP model with two hyper-parameters is difficult. Fair model compassion can be performed using Bayesian evidence analysis based on parameter posterior distribution probabilities, but once again, this requires informative parameter priors of the GP, which we do not have. Finally, as  \cite{2017A&A...597A.128K} found, the temporal baseline of the observations must be at least ten times longer than the true DRW decorrelation timescale. However, the characteristic corelation time scale of our DRW model is $\approx$ 660$_{-80}^{+1800}$ days, while we have a temporal baseline of only $\sim$ 15 years, which is insufficient for conclusive DRW results.
 
Therefore, we find that the DRW model is likely inappropriate to describe our data despite the relatively good fit properties. We conclude that more credibility to the DRW possibility of QSO B1312+7837 is possible only if more optical photometry data are collected, but that would require a few decades of observations.

\subsection{Binary SMBH model -- physical parameters}

The spectrum allows us to estimate of the two SMBH masses from for the 
width of the broad C{\sc iv} 1549.06\,\AA\ component and the continuum 
luminosity at 1350\,\AA\ from \citet[][Eqs. 2 and 4]{2006ApJ...641..689V}. 
Two scenarios must be considered here. First, if the separation between 
the two SMBHs is wide, so each has its own accretion disk, and the orbital 
motion of the SMBHs around their common center of masses does not contribute 
significantly to the width of the emission lines, then the final mass will 
be the luminosity weighted average of the masses of the two SMBHs. 
Second, if the SMBHs are close in, and are immersed into a single accretion 
disk, we will obtain from the width of the broad emission lines an upper 
limit to the combined mass of the two SMBHs, because the motion of the 
binary SMBHs is expected to produce additional widening of the emission 
lines. We can not distinguish between the two options and can only 
conclude that the combined mass of the two SMBHs must be lower than twice 
the obtained estimate.

Given the resolution of our spectra the deblending of the narrow and the 
broad C{\sc iv} components gives uncertain results. We performed it with 
the IRAF task {\it splot} and obtained widths of 
2700$\pm$200\,km\,s$^{-1}$ and 9700$\pm$600\,km\,s$^{-1}$, respectively.
The relation of \citet{{2006ApJ...641..689V}} yields an estimate of 
$log$(M$_{BH}$/M$_\odot$)$\sim$8.1$\pm$0.1 which, as discussed above, 
implies that the combined mass of the two SMBHs in the system can not 
exceed twice this value.

Assuming a circular orbit, for the derived rest frame period, a mass 
ratio of the 
two black holes between 0.5 and 1, and masses in the generously wide 
range $log$(M$_{BH}$/M$_\odot$)$\sim$8--8.6 we obtain orbital velocity
for the primary SBMH in the range of 3600--8600\,km\,s$^{-1}$ and for 
the secondary companion -- of 5500--11500\,km\,s$^{-1}$, following 
\citet{2015Natur.525..351D}. This is $\sim$1-4\,\% of the speed of 
light, sufficient to cause relativistic boost. 
 
The flux change caused by Doppler boosting of a power low spectrum 
F$_\nu\sim\nu^{\alpha}$ is 
F$^{max}$/F$^{min}$= ((1+$\beta$sin$i$)/(1-$\beta$sin$i$))$^{(3-\alpha)}$
\citep[e.g.,][]{1987ApJ...319..416P,2010A&A...516A..18D}. From our 
spectrum, after a correction for Milky Way reddening using the 
extinction law of \citet{1989ApJ...345..245C} and A$_V\sim0.09$\,mag 
from \citet{2011ApJ...737..103S}, we measured an average slope 
$\alpha$=$-$0.24$\pm$0.05 over $\lambda\sim$1400--2600\,\AA.
The slope error is tentatively adopted conservative value: the formal 
fitting error is typically 0.02-0.03, but the slope varies within 
$\sim$0.05 depending on the wavelength range and the masking of the 
emission lines. We carried out the same analysis of the archival 
spectrum from \citet{1999A&AS..134..483H} and obtained similar slope.
To explain the observed F$^{max}$/F$^{min}\sim$0.2 this implies that 
the line of sight velocity needs to be 
{\it v}\,$sin$\,$i$\,$\sim$8550\,km\,s$^{-1}$, at the limit of the 
orbital velocity for the primary black hole, but well within the 
range of orbital velocity for the secondary. 

For the mass range and the derived period the Kepler's third law 
yields a semi-major axis of 0.008-0.012\,pc which is significantly 
larger than the innermost marginally stable circular orbit for SMBHs 
within the considered mass range: r$_{IRSO}\sim$10$^{-4}$--10$^{-5}$
\citep[for a SMBH with spin $a$=0 or even six times smaller for a SMBH 
with spin $a$=1;][]{2010ApJ...714..404T}.

To place QSO B1312+7837 in broader context we compare the physical 
parameters we derived for this object with the parameters of quasars 
in the \citet{2010ApJ...721.1014M} sample. For apparent $i$ magnitude 
16.052$\pm$0.005\,mag \citep{2016arXiv161205560C} and adopting a 
cosmological model with H$_0$=70.0\,km\,s$^{-1}$\,Mpc$^{-1}$ and
$\Omega_M$=0.3, yielding a distance modulus of 
$\sim$45.95\,mag\footnote{\url{https://cosmocalc.icrar.org/}} we obtain 
M$_i\sim$-29.9\,mag plasing QSO B1312+7837 at the brightest end of 
the quasar locus -- this is expected for quasars at higher redshift 
but it also hints that the true black hole mass of QSO B1312+7837 is 
more likely to be closer to the upper end of our mass range
\citep[fig. 12 in][]{2010ApJ...721.1014M}.

\section{Discussion and Conclusions}\label{sec:discussion}

The binary SMBHs evade detection for a number of reasons: because the 
velocity difference is smaller than the intrinsic width of the emission 
lines or because the offset nuclei are too close to be resolved with the 
existing instrumentation etc. Other techniques require competitive 
observing time at the few X-ray missions or moderately high resolution 
high multiplexity vast spectroscopic surveys. On the other hand the new 
or soon to enter operation all sky variability monitoring projects like 
Pan-STARRS and LSST/VRO \citep{2002SPIE.4836..154K,2019ApJ...873..111I} 
will make the demographic studies of binary SMBHs more accessible.

The masses of quasars were first estimated with arguments related with 
their bolometric luminosities \citep{1982MNRAS.200..115S}. The 
reverberation mapping \citep{1982ApJ...255..419B} provided more accurate
estimates and the scalling relations with the bulge properties
\citep{2000ApJ...539L...9F,2000ApJ...539L..13G} allowed for studies of
SMBH demographics. The vast majority of SMBHs have masses in the range
$log$(M$_{BHs}$/M$_\odot$)$\sim$7-9 \citep{2004MNRAS.351..169M}, with 
low and high record holders of $log$(M$_{BHs}$/M$_\odot$)$\sim$5 and 11 
\citep{2015ApJ...809L..14B,2014ApJ...795L..31L}. Our upper mass limit 
is well within this range.

The estimated orbital elements make QSO B1312+7837 similar to PG 1302-102, 
a well known sub-pc separation quasar with at $z$=0.3 that shows sinusoidal 
variations in the optical \citep{2015Natur.518...74G,2015Natur.525..351D}. 
However, straightforward interpretation of the observed period and all 
other related properties is hampered by the unknown mass ratio of the two 
SMBH components and the decoupling between the SMBH orbital period and the 
strongest periodicity in the accretion rate onto the SMBHs -- 
\citet{2015MNRAS.452.2540D} pointed out that the latter may correspond to 
the orbital period of accreted gas at a distance of a few binary 
separations. Therefore, the binary SMBH might be much tighter than 
suggested by the 2-yr rest-frame period, and even be in the 
gravitational-wave dominated orbital decay regimen. X-ray spectroscopy of 
features formed in the innermost region of the accretion disk may offer 
the only opportunity to probe directly the SMBH orbit, their mass ratio
and the dominant orbital decay mechanism. Finally, if the Doppler
boosting account for the flux modulation, the calculated orbital 
velocities imply that the majority of the emission probably originates 
around the secondary component, because the observed amplitude can only 
be accounted for with velocities at the limit of the velocity range 
for the primary.

We note that the sinusoidal light curve can be explained by other models,
including the presence of a hot spot in the inner accretion disk or a 
warp in the disk itself \citep{2015Natur.518...74G}. It is still possible 
that periodic variations in the light curves of 
quasars could be a manifestation of correlated "red-noise" stochastic 
systematics. For example, \citet{2016MNRAS.461.3145V} also studied the 
light curve of PG 1302--102, and found that stochastic ``red-noise''
processes are likely preferred over a sinusoidal variation. 
For precise stochastic process analyses of QSO B1312+7837 we need longer baseline observations 
with high cadence to provide accurate results from the models.

Summarizing, we found a variation with a probable period of 2214$\pm$12\,d in 
the apparent brightness of the $z$=2.0 QSO B1312+7837. This modulation 
can be described with a binary SMBH with a combined upper mass limit of
$log$(M$_{BHs}$/M$_\odot$)$\sim$8.4$\pm$0.1 -- a value that places this 
object within the typical SMBH mass range. We argue that the advent of 
all-sky synoptic surveys will soon allow to carry out studies of the 
SMBHs demographics.


\section*{Acknowledgements}

We thank to Zhang-Liang Xie for very useful discussion, and for confirming 
our DRW results using {\tt JAVELIN} code \citep{2013ApJ...765...106Z}.
We acknowledge support by Bulgarian NSF under grant DN18-10/2017 and 
National RI Roadmap Projects DO1-277/16.12.2019 and DO1-268/16.12.2019 
of the Ministry of Education and Science of the Republic of Bulgaria.
We report to study RFBR project \textnumero\,19-02-00432.  
T.T.\ acknowledge support by the DFG Research Unit FOR~2544 {\it Blue  Planets around Red Stars}.
Observations with the SAO RAS telescopes are supported by the Ministry of
Science and Higher Education of the Russian Federation (including
agreement No05.619.21.0016, project ID RFMEFI61919X0016). Optical
spectroscopy of the sources was performed as part of the government
contract of the SAO RAS approved by Ministry of Science and Higher
Education of the Russian Federation.
The authors thank the anonymous referee for the useful comments.

\section*{Data availability}
The data underlying this article are available in the article and in its online supplementary material.





\bibliographystyle{mnras}

\begin{thebibliography}{99}
\bibitem[\protect\citeauthoryear{Afanasiev \& Moiseev}{2005}]{2005AstL...31..194A} Afanasiev V.~L., Moiseev A.~V., 2005, AstL, 31, 194. doi:10.1134/1.1883351
\bibitem[\protect\citeauthoryear{Bachev, Strigachev, \& Semkov}{2005}]{2005MNRAS.358..774B} Bachev R., Strigachev A., Semkov E., 2005, MNRAS, 358, 774. doi:10.1111/j.1365-2966.2005.08708.x
\bibitem[\protect\citeauthoryear{Baldassare et al.}{2015}]{2015ApJ...809L..14B} Baldassare V.~F., Reines A.~E., Gallo E., Greene J.~E., 2015, ApJL, 809, L14. doi:10.1088/2041-8205/809/1/L14
\bibitem[\protect\citeauthoryear{Baluev}{2009}]{2009MNRAS.393..969B} Baluev R.~V., 2009, MNRAS, 393, 969. doi:10.1111/j.1365-2966.2008.14217.x
\bibitem[\protect\citeauthoryear{Begelman, Blandford, \& Rees}{1980}]{1980Natur.287..307B} Begelman M.~C., Blandford R.~D., Rees M.~J., 1980, Natur, 287, 307. doi:10.1038/287307a0
\bibitem[\protect\citeauthoryear{Blandford \& McKee}{1982}]{1982ApJ...255..419B} Blandford R.~D., McKee C.~F., 1982, ApJ, 255, 419. doi:10.1086/159843
\bibitem[\protect\citeauthoryear{C{\'a}ceres et al.}{2009}]{2009A&A...507..481C} C{\'a}ceres C., Ivanov V.~D., Minniti D., Naef D., Melo C., Mason E., Selman F., et al., 2009, \aap, 507, 481. doi:10.1051/0004-6361/200810908
\bibitem[\protect\citeauthoryear{C{\'a}ceres et al.}{2011}]{2011A&A...530A...5C} C{\'a}ceres C., Ivanov V.~D., Minniti D., Burrows A., Selman F., Melo C., Naef D., et al., 2011, \aap, 530, A5. doi:10.1051/0004-6361/201016231
\bibitem[\protect\citeauthoryear{Cardelli, Clayton, \& Mathis}{1989}]{1989ApJ...345..245C} Cardelli J.~A., Clayton G.~C., Mathis J.~S., 1989, ApJ, 345, 245. doi:10.1086/167900
\bibitem[\protect\citeauthoryear{Chambers et al.}{2016}]{2016arXiv161205560C} Chambers K.~C., Magnier E.~A., Metcalfe N., Flewelling H.~A., Huber M.~E., Waters C.~Z., Denneau L., et al., 2016, arXiv, arXiv:1612.05560
\bibitem[\protect\citeauthoryear{Charisi et al.}{2016}]{2016MNRAS.463.2145C} Charisi M., Bartos I., Haiman Z., Price-Whelan A.~M., Graham M.~J., Bellm E.~C., Laher R.~R., et al., 2016, MNRAS, 463, 2145. doi:10.1093/mnras/stw1838
\bibitem[\protect\citeauthoryear{Dubus, Cerutti, \& Henri}{2010}]{2010A&A...516A..18D} Dubus G., Cerutti B., Henri G., 2010, A\&A, 516, A18. doi:10.1051/0004-6361/201014023
\bibitem[\protect\citeauthoryear{D'Orazio, Haiman, \& MacFadyen}{2013}]{2013MNRAS.436.2997D} D'Orazio D.~J., Haiman Z., MacFadyen A., 2013, MNRAS, 436, 2997. doi:10.1093/mnras/stt1787
\bibitem[\protect\citeauthoryear{D'Orazio et al.}{2015a}]{2015MNRAS.452.2540D} D'Orazio D.~J., Haiman Z., Duffell P., Farris B.~D., MacFadyen A.~I., 2015a, MNRAS, 452, 2540. doi:10.1093/mnras/stv1457
\bibitem[\protect\citeauthoryear{D'Orazio, Haiman, \& Schiminovich}{2015b}]{2015Natur.525..351D} D'Orazio D.~J., Haiman Z., Schiminovich D., 2015b, Natur, 525, 351. doi:10.1038/nature15262
\bibitem[\protect\citeauthoryear{D'Orazio \& Loeb}{2018}]{2018ApJ...863..185D} D'Orazio D.~J., Loeb A., 2018, ApJ, 863, 185. doi:10.3847/1538-4357/aad413
\bibitem[\protect\citeauthoryear{Eracleous \& Halpern}{1994}]{1994ApJS...90....1E} Eracleous M., Halpern J.~P., 1994, ApJS, 90, 1. doi:10.1086/191856
\bibitem[\protect\citeauthoryear{Fan et al.}{1998}]{1998ApJ...507..173F} Fan J.~H., Xie G.~Z., Pecontal E., Pecontal A., Copin Y., 1998, ApJ, 507, 173. doi:10.1086/306301
\bibitem[\protect\citeauthoryear{Ferrarese \& Merritt}{2000}]{2000ApJ...539L...9F} Ferrarese L., Merritt D., 2000, ApJL, 539, L9. doi:10.1086/312838
\bibitem[\protect\citeauthoryear{Foreman-Mackey et al.}{2013}]{2013PASP..125..306F} Foreman-Mackey D., Hogg D.~W., Lang D., Goodman J., 2013, PASP, 125, 306. doi:10.1086/670067
\bibitem[\protect\citeauthoryear{Foreman-Mackey et al.}{2017}]{2017AJ....154..220F} Foreman-Mackey D., Agol E., Ambikasaran S., Angus R., 2017, AJ, 154, 220. doi:10.3847/1538-3881/aa9332
\bibitem[\protect\citeauthoryear{Gebhardt et al.}{2000}]{2000ApJ...539L..13G} Gebhardt K., Bender R., Bower G., Dressler A., Faber S.~M., Filippenko A.~V., Green R., et al., 2000, ApJL, 539, L13. doi:10.1086/312840
\bibitem[\protect\citeauthoryear{Goodman \& Weare}{2010}]{2010CAMCS...5...65G} Goodman J., Weare J., 2010, CAMCS, 5, 65. doi:10.2140/camcos.2010.5.65
\bibitem[\protect\citeauthoryear{Graham et al.}{2015}]{2015MNRAS.453.1562G} Graham M.~J., Djorgovski S.~G., Stern D., Drake A.~J., Mahabal A.~A., Donalek C., Glikman E., et al., 2015, MNRAS, 453, 1562. doi:10.1093/mnras/stv1726
\bibitem[\protect\citeauthoryear{Graham et al.}{2015}]{2015Natur.518...74G} Graham M.~J., Djorgovski S.~G., Stern D., Glikman E., Drake A.~J., Mahabal A.~A., Donalek C., et al., 2015, Natur, 518, 74. doi:10.1038/nature14143
\bibitem[\protect\citeauthoryear{Hagen, Engels, \& Reimers}{1999}]{1999A&AS..134..483H} Hagen H.-J., Engels D., Reimers D., 1999, A\&AS, 134, 483. doi:10.1051/aas:1999442
\bibitem[\protect\citeauthoryear{Halpern \& Filippenko}{1988}]{1988Natur.331...46H} Halpern J.~P., Filippenko A.~V., 1988, Natur, 331, 46. doi:10.1038/331046a0
\bibitem[\protect\citeauthoryear{Henden et al.}{2016}]{2016yCat.2336....OH} Henden A.~A., Templeton M., Terrell D., Smith T.~C., Levine S., Welch D., 2016, AAVSO Photometric All Sky Survey (APASS) DR9, II/336/apass9
\bibitem[\protect\citeauthoryear{Hills}{1975}]{1975Natur.254..295H} Hills J.~G., 1975, Natur, 254, 295. doi:10.1038/254295a0
\bibitem[\protect\citeauthoryear{Ivanov et al.}{2016}]{2016A&A...588A..93I} Ivanov V.~D., Cioni M.-R.~L., Bekki K., de Grijs R., Emerson J., Gibson B.~K., Kamath D., et al., 2016, \aap, 588, A93. doi:10.1051/0004-6361/201527398
\bibitem[\protect\citeauthoryear{Ivezi{\'c} et al.}{2019}]{2019ApJ...873..111I} Ivezi{\'c} {\v{Z}}., Kahn S.~M., Tyson J.~A., Abel B., Acosta E., Allsman R., Alonso D., et al., 2019, ApJ, 873, 111. doi:10.3847/1538-4357/ab042c
\bibitem[\protect\citeauthoryear{Jockers et al.}{2000}]{2000KFNTS...3...13J} Jockers K., Credner T., Bonev T., Kisele V.~N., Korsun P., Kulyk I., Rosenbush V., et al., 2000, KFNTS, 3, 13.
\bibitem[\protect\citeauthoryear{L{\'o}pez-Cruz et al.}{2014}]{2014ApJ...795L..31L} L{\'o}pez-Cruz O., A{\~n}orve C., Birkinshaw M., Worrall D.~M., Ibarra-Medel H.~J., Barkhouse W.~A., Torres-Papaqui J.~P., et al., 2014, ApJL, 795, L31. doi:10.1088/2041-8205/795/2/L31
\bibitem[\protect\citeauthoryear{Kaiser et al.}{2002}]{2002SPIE.4836..154K} Kaiser N., Aussel H., Burke B.~E., Boesgaard H., Chambers K., Chun M.~R., Heasley J.~N., et al., 2002, SPIE, 4836, 154. doi:10.1117/12.457365
\bibitem[\protect\citeauthoryear{Kelley et al.}{2019}]{2019MNRAS.485.1579K} Kelley L.~Z., Haiman Z., Sesana A., Hernquist L., 2019, MNRAS, 485, 1579. doi:10.1093/mnras/stz150
\bibitem[\protect\citeauthoryear{Kelly et al.}{2009}]{2009ApJ...698..895K} Kelly B.~C., Bechtold J., Siemiginowska A., 2009, ApJ, 698, 895. doi:10.1088/0004-637X/698/1/895
\bibitem[\protect\citeauthoryear{Komossa et al.}{2003}]{2003ApJ...582L..15K} Komossa S., Burwitz V., Hasinger G., Predehl P., Kaastra J.~S., Ikebe Y., 2003, ApJL, 582, L15. doi:10.1086/346145
\bibitem[\protect\citeauthoryear{Komossa}{2006}]{2006MmSAI..77..733K} Komossa S., 2006, MmSAI, 77, 733
\bibitem[\protect\citeauthoryear{Kozlowski et al.}{2010}]{2010ApJ...708..927K} Kozlowski S., Kochanek C., et al., 2010, ApJ, 708, 927. doi:10.1088/0004-637X/708/2/927
\bibitem[\protect\citeauthoryear{Kozlowski}{2017}]{2017A&A...597A.128K} Kozlowski S., 2017, \aap, 597, A128. doi:10.1051/0004-6361/201629890
\bibitem[\protect\citeauthoryear{Lehto \& Valtonen}{1996}]{1996ApJ...460..207L} Lehto H.~J., Valtonen M.~J., 1996, ApJ, 460, 207. doi:10.1086/176962
\bibitem[\protect\citeauthoryear{Liu et al.}{2016}]{2016ApJ...833....6L} Liu T., Gezari S., Burgett W., Chambers K., Draper P., Hodapp K., Huber M., et al., 2016, ApJ, 833, 6. doi:10.3847/0004-637X/833/1/6
\bibitem[\protect\citeauthoryear{MacLeod et al.}{2010}]{2010ApJ...721.1014M} MacLeod, C.~L., Ivezic, Z., et al., 2010, ApJ, 721, 1014. doi:10.1088/0004-637X/721/2/1014
\bibitem[\protect\citeauthoryear{Marconi et al.}{2004}]{2004MNRAS.351..169M} Marconi A., Risaliti G., Gilli R., Hunt L.~K., Maiolino R., Salvati M., 2004, MNRAS, 351, 169. doi:10.1111/j.1365-2966.2004.07765.x
\bibitem[\protect\citeauthoryear{Mickaelian et al.}{1999}]{1999Afz....42....5M} Mickaelian A.~M., Gon{\c{c}}alves A.~C., V{\'e}ron-Cetty M.~P., V{\'e}ron P., 1999, Afz, 42, 5
\bibitem[\protect\citeauthoryear{Nelder \& Mead}{1965}]{NelderMead65} Nelder J.~A. \& Mead R., 1965, The Computer Journal, 7, 308; doi:10.1093/comjnl/7.4.308
\bibitem[\protect\citeauthoryear{Ovcharov et al.}{2008}]{2008MNRAS.386..819O} Ovcharov E.~P., Nedialkov P.~L., Valcheva A.~T., Ivanov V.~D., Tikhonov N.~A., Stanev I.~S., Kostov A.~B., et al., 2008, MNRAS, 386, 819. doi:10.1111/j.1365-2966.2008.12990.x
\bibitem[\protect\citeauthoryear{Pelling et al.}{1987}]{1987ApJ...319..416P} Pelling R.~M., Paciesas W.~S., Peterson L.~E., Makishima K., Oda M., Ogawara Y., Miyamoto S., 1987, ApJ, 319, 416. doi:10.1086/165466
\bibitem[\protect\citeauthoryear{Rubinur et al.}{2021}]{2021MNRAS.500.3908R} Rubinur K., Kharb P., Das M., Rahna P.~T., Honey M., Paswan A., Vaddi S., et al., 2021, MNRAS, 500, 3908. doi:10.1093/mnras/staa3375
\bibitem[\protect\citeauthoryear{Schlafly \& Finkbeiner}{2011}]{2011ApJ...737..103S} Schlafly E.~F., Finkbeiner D.~P., 2011, ApJ, 737, 103. doi:10.1088/0004-637X/737/2/103
\bibitem[\protect\citeauthoryear{Severgnini et al.}{2021}]{2021A&A...646A.153S} Severgnini P., Braito V., Cicone C., Saracco P., Vignali C., Serafinelli R., Della Ceca R., et al., 2021, \aap, 646, A153. doi: 10.1051/0004-6361/202039576
\bibitem[\protect\citeauthoryear{Sillanpaa et al.}{1988}]{1988ApJ...325..628S} Sillanpaa A., Haarala S., Valtonen M.~J., Sundelius B., Byrd G.~G., 1988, ApJ, 325, 628. doi:10.1086/166033
\bibitem[\protect\citeauthoryear{Smith et al.}{2010}]{2010ApJ...716..866S} Smith K.~L., Shields G.~A., Bonning E.~W., McMullen C.~C., Rosario D.~J., Salviander S., 2010, ApJ, 716, 866. doi:10.1088/0004-637X/716/1/866
\bibitem[\protect\citeauthoryear{Soltan}{1982}]{1982MNRAS.200..115S} Soltan A., 1982, MNRAS, 200, 115. doi:10.1093/mnras/200.1.115
\bibitem[\protect\citeauthoryear{Souchay et al.}{2015}]{2015A&A...583A..75S} Souchay J., Andrei A.~H., Barache C., Kalewicz T., Gattano C., Coelho B., Taris F., et al., 2015, A\&A, 583, A75. doi:10.1051/0004-6361/201526092
\bibitem[\protect\citeauthoryear{Stetson}{2000}]{2000PASP..112..925S} Stetson P.~B., 2000, PASP, 112, 925. doi:10.1086/316595
\bibitem[\protect\citeauthoryear{Tanaka \& Menou}{2010}]{2010ApJ...714..404T} Tanaka T., Menou K., 2010, ApJ, 714, 404. doi:10.1088/0004-637X/714/1/404
\bibitem[\protect\citeauthoryear{Tody}{1986}]{1986SPIE..627..733T} Tody D., 1986, SPIE, 627, 733. doi:10.1117/12.968154
\bibitem[\protect\citeauthoryear{Tody}{1993}]{1993ASPC...52..173T} Tody D., 1993, ASPC, 52, 173
\bibitem[\protect\citeauthoryear{Trifonov}{2019}]{2019ascl.soft06004T} Trifonov T., 2019, ascl.soft. ascl:1906.004
\bibitem[\protect\citeauthoryear{Tytler et al.}{2004}]{2004AJ....128.1058T} Tytler D., O'Meara J.~M., Suzuki N., et al., 2004, AJ, 128, 1058. doi:10.1086/423293
\bibitem[\protect\citeauthoryear{Vanden Berk et al.}{2001}]{2001AJ....122..549V} Vanden Berk D.~E., Richards G.~T., Bauer A., et al., 2001, AJ, 122, 549. doi:10.1086/321167
\bibitem[\protect\citeauthoryear{Vaughan et al.}{2016}]{2016MNRAS.461.3145V} Vaughan S., Uttley P., Markowitz A.~G., et al., 2016, MNRAS, 461, 3145. doi:10.1093/mnras/stw1412
\bibitem[\protect\citeauthoryear{V{\'e}ron-Cetty \& V{\'e}ron}{2006}]{2006A&A...455..773V} V{\'e}ron-Cetty M.-P., V{\'e}ron P., 2006, A\&A, 455, 773. doi:10.1051/0004-6361:20065177
\bibitem[\protect\citeauthoryear{Vestergaard \& Peterson}{2006}]{2006ApJ...641..689V} Vestergaard M., Peterson B.~M., 2006, ApJ, 641, 689. doi:10.1086/500572
\bibitem[\protect\citeauthoryear{Ward et al.}{2021}]{2021ApJ...913..102W} Ward C., Gezari S., Frederick S., Hammerstein E., Nugent P., van Velzen S., Drake A., et al., 2021, ApJ, 913, 102. doi: 10.3847/1538-4357/abf246
\bibitem[\protect\citeauthoryear{Wolfe \& Burbidge}{1970}]{1970ApJ...161..419W} Wolfe A.~M., Burbidge G.~R., 1970, ApJ, 161, 419. doi:10.1086/150549
\bibitem[\protect\citeauthoryear{Zechmeister \& K\"{u}rster}{2009}]{2009A&A...496..577Z} Zechmeister M., K\"{u}rster M, 2009, A\&A, 496, 577. doi:10.1051/0004-6361:200811296
\bibitem[\protect\citeauthoryear{Zu et al.}{2013}]{2013ApJ...765...106Z} Zu Y., Kochanek, C.~S., Kozlowski, S., Udalski, A., 2013, ApJ, 765, 106. doi:10.1088/0004-637X/765/2/106
\end{thebibliography}



\bsp	
\label{lastpage}
\end{document}